# Navigability evaluation of complex networks by greedy routing efficiency

Alessandro Muscoloni[1] and Carlo Vittorio Cannistraci[1,2,*]

[1]Biomedical Cybernetics Group, Biotechnology Center (BIOTEC), Center for Molecular and Cellular Bioengineering (CMCB), Center for Systems Biology Dresden (CSBD), Department of Physics, Technische Universität Dresden, Tatzberg 47/49, 01307 Dresden, Germany
[2]Brain bio-inspired computing (BBC) lab, IRCCS Centro Neurolesi "Bonino Pulejo", Messina, Italy
*Corresponding author: Carlo Vittorio Cannistraci (kalokagathos.agon@gmail.com)

**Abstract**

Network navigability is a key feature of complex networked systems. For a network embedded in a geometrical space, maximization of greedy routing (GR) measures based on the node geometrical coordinates can ensure efficient greedy navigability. In PNAS, Seguin et al. (PNAS 2018, vol. 115, no. 24) define a measure for quantifying the efficiency of brain network navigability in the Euclidean space, referred to as the *efficiency ratio*, whose formula exactly coincides with the *GR-score (GR-efficiency)* previously published by Muscoloni et al. (Nature Communications 2017, vol. 8, no. 1615). In this Letter, we point out potential flaws in the study of Seguin et al. regarding the discussion of the GR evaluation. In particular, we revise the concept of GR navigability, together with a careful discussion of the advantage offered by the new proposed GR-efficiency measure in comparison to the main measures previously adopted in literature. Finally, we clarify and standardize the GR-efficiency terminology in order to simplify and facilitate the discussion in future studies.

## Letter

An important characteristic of any network is its navigability. The navigability of a network embedded in a geometrical space is associated with the fact that greedy routing (GR) performed using the node geometrical coordinates is efficient [1]–[5]. In PNAS, Seguin et al. [6] define a measure for quantifying the efficiency of brain network navigability in the Euclidean space, referred to as the *efficiency ratio*, whose formula exactly coincides with the *GR-score* previously introduced by Muscoloni et al. [7]. However, in contrast to Muscoloni et al. [7], the study of Seguin et al. [6] (i) proposes the *efficiency ratio* without referring to GR at all, (ii) provides no rationale for the introduction of the *efficiency ratio* with respect to the two main GR measures already adopted in the literature, and (iii) does not discuss any advantage of the measure, nor does it explain that the *efficiency ratio* in fact integrates the two former measures. Therefore, we feel that, for the sake of completeness, the following points should be clarified. GR navigability implies that, for each pair of nodes, a packet is sent from the source to the destination, and that each node knows only the address (coordinates) of its neighbors and the address of the destination, which is written in the packet. At each hop the packet is greedily forwarded from the current node to the neighbor, which is closest to the destination, and is dropped when a loop is detected [1]–[5]. Before the study of Muscoloni et al. [7], two measures were primarily adopted to evaluate the GR: (i) the *success ratio*, defined as the percentage of successful paths, and (ii) the *stretch*, defined as the average ratio of GR path length over shortest path length for successful GR paths [2]–[5]. In 2017, Muscoloni et al. [7] introduced a new measure for evaluating the efficiency of GR, named the GR-score, and defined this as the average ratio of shortest path length over GR path length (where the path length of unsuccessful GR paths is infinite). This measure allows assessment of GR navigability with a unique score that integrates both the concepts of success ratio and stretch: a path is assigned a GR-score of 0 if unsuccessful (worst case), a GR-score of ]0,1[ if successful with a stretch greater than 1, and a GR-score of 1 if successful with a stretch 1 (best case). As shown in Fig. 1, the introduction of the GR-score provides a unique solution when success ratio and stretch suggest conflicting results.

Finally, clarification of the terminology is necessary. Seguin et al. [6] refer to the routing model adopted simply as "navigation," whereas a more appropriate term is "greedy navigation" or "greedy routing," since "navigation" can be interpreted more generally as any forwarding protocol, not necessarily greedy, unless explicitly stated. Similarly, the term "efficiency ratio" may be misleading, since it appears as a general concept and there is no indication that it is derived from a very specific navigation protocol, the GR. We therefore suggest adoption of the term "GR-score" or, alternatively, "GR efficiency" in further studies.

## Acknowledgments

We thank Gary Jennings for proofreading the text.

**References**

[1]    J. M. Kleinberg, "Navigation in a small world," *Nature*, 2000.

[2]    M. Boguñá, D. Krioukov, and K. C. Claffy, "Navigability of complex networks," *Nat. Phys.*, vol. 5, no. 1, pp. 74–80, 2008.

[3]    M. Boguñá, F. Papadopoulos, and D. Krioukov, "Sustaining the Internet with Hyperbolic Mapping," *Nat. Commun.*, vol. 1, no. 6, pp. 1–8, 2010.

[4]    F. Papadopoulos, C. Psomas, and D. Krioukov, "Network mapping by replaying hyperbolic growth," *IEEE/ACM Trans. Netw.*, vol. 23, no. 1, pp. 198–211, 2015.

[5]    F. Papadopoulos, R. Aldecoa, and D. Krioukov, "Network Geometry Inference using Common Neighbors," *Phys. Rev. E*, vol. 92, no. 2, p. 022807, 2015.

[6]    C. Seguin, M. P. van den Heuvel, and A. Zalesky, "Navigation of brain networks," *Proc. Natl. Acad. Sci. U. S. A.*, no. 11, 2018.

[7]    A. Muscoloni, J. M. Thomas, S. Ciucci, G. Bianconi, and C. V. Cannistraci, "Machine learning meets complex networks via coalescent embedding in the hyperbolic space," *Nat. Commun.*, vol. 8, 2017.

[8]    A. Muscoloni and C. V. Cannistraci, "A nonuniform popularity-similarity optimization (nPSO) model to efficiently generate realistic complex networks with communities," *New J. Phys.*, vol. 20, 2018.

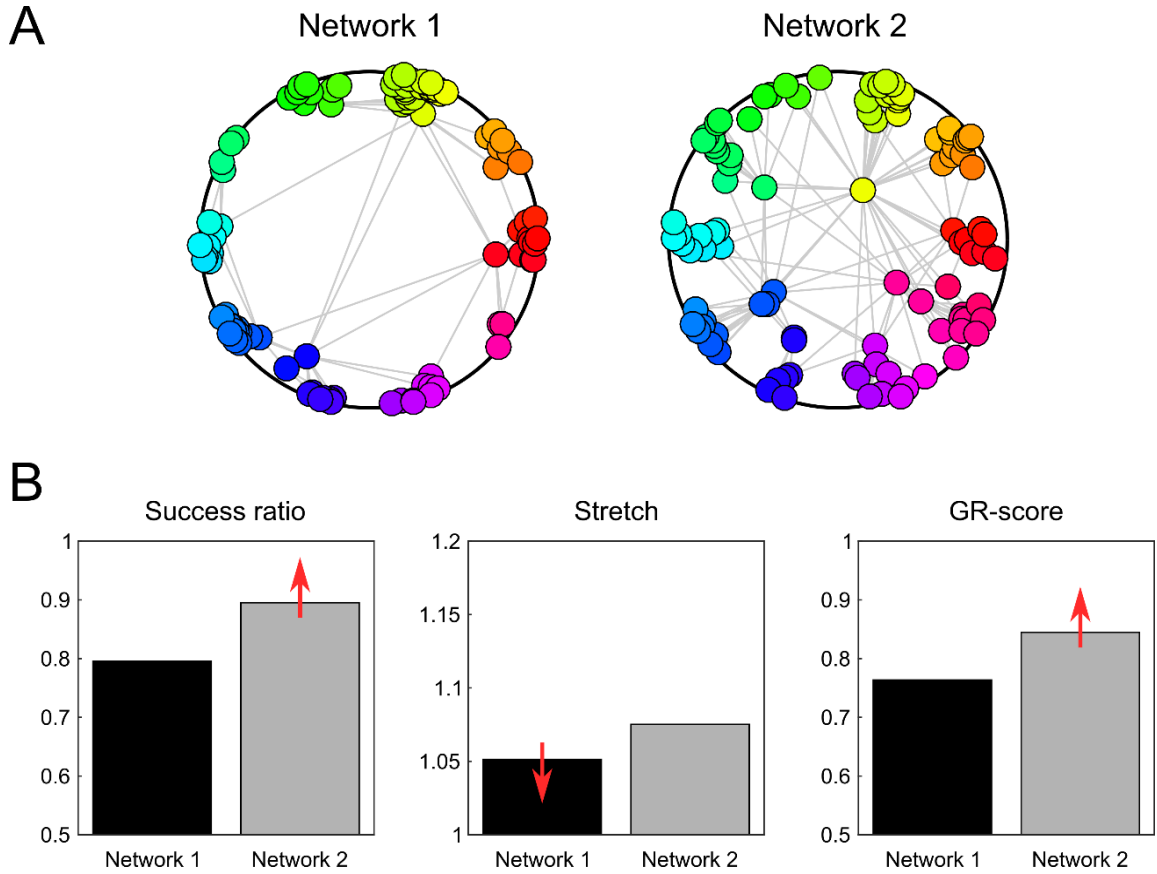

**Fig. 1.** **(A)** Two artificial networks generated in the hyperbolic space using the nonuniform popularity-similarity optimization (nPSO) model [8], with the following parameters: network 1 $N = 100$, $m = 2$, $T = 0.1$, $\gamma = 5$ and $C = 10$; network 2 $N = 100$, $m = 2$, $T = 0.3$, $\gamma = 2.5$ and $C = 10$; $N$ is the network size, $m$ is approximately half of the average node degree, $T$ is the network temperature (inversely related to the clustering), $\gamma$ is the exponent of the scale-free degree distribution, and $C$ is the number of communities. **(B)** Bar plots comparing the success ratio, stretch, and GR-score for the two networks; the red arrow indicates which network has a better result for each evaluation measure. Due to the high $\gamma$, network 1 is lacking hubs that act as bridges in the navigation between nodes far from each other, and therefore it has a lower success ratio than network 2. However, the successful GR paths of network 1 have a lower stretch than the more numerous ones of network 2, since they are mostly local paths and the higher clustering (lower temperature) facilitates the navigation between nearby nodes. The GR-score provides a unique solution to the conflicting results of the other two measures, suggesting that network 2 has higher navigability, which is reasonable since the success ratio in network 2 is ~12% higher than in network 1, whereas the stretch decrement of network 1 on network 2 is ~2%.